\documentclass{appolb}
\usepackage{epsfig}
\begin{document}
\title{Signatures of the chiral critical endpoint of QCD in heavy-ion collisions: the role of finite-size effects%
\thanks{Presented at Excited QCD 2011 (Les Houches, France).}%
%%%% upto 6 pages
}
\author{Let\'\i cia F. Palhares
\address{Institut de Physique Th\'eorique, CEA-Saclay,
% 91191 Gif-sur-Yvette Cedex,
 France}
\address{
Instituto de F\'\i sica, Universidade Federal do Rio de Janeiro,
%Caixa Postal 68528, Rio de Janeiro, RJ 21941-972, 
Brazil}
\and
Eduardo S. Fraga
\address{Instituto de F\'\i sica, Universidade Federal do Rio de Janeiro,
% Caixa Postal 68528, Rio de Janeiro, RJ 21941-972,
 Brazil}
}
\maketitle
\begin{abstract}
We briefly discuss the status of the signatures of the QCD critical endpoint in heavy-ion collisions concentrating on the role played by the finiteness of the medium created in such experiments.
\end{abstract}
  
\section{Introduction}

After more than a decade of experiments colliding heavy-ion nuclei to investigate the properties of matter in extreme conditions of temperature and density, a clear signal of collectivity at the partonic level observed at RHIC-BNL together with other indications point to the formation of a new state of matter, even though its properties seem to be more intricate than those of the originally predicted quark-gluon plasma \cite{Adams:2005dq}. Current and near-future colliders stretch the currently probed frontiers in two directions. The higher energies attained at the LHC provide the closest conditions to the primordial universe plasma and the rich statistics will allow for better studies of fluctuations and the properties of the medium. In the other direction, different experimental programs are dedicated to probe ultra-dense media with higher baryon-antibaryon asymmetry: the on-going Beam Energy Scan (BES) program \cite{starbes} at RHIC-BNL and the future colliders at FAIR-GSI and NICA-JINR. This region of the phase diagram is expected to contain the richer part of the phase structure of strong interactions. Unfortunately, it is still obscure from the point of view of theoretical first principle calculations, while a plethora of different effective models and approximate nonperturbative approaches predict different phases and phase boundaries \cite{Stephanov:2004wx}. Any experimental guidance would be invaluable in this quest for understanding the phase structure of QCD at high energies and that is the opportunity provided by this set of experiments in the next years.

In particular, the establishment of the existence and location of the critical endpoint of QCD (CEP) is a crucial step in the mapping of the phase diagram. The investigation of the experimental signatures of the CEP in the specific setup of heavy-ion collisions (HICs) is therefore of great importance. In this paper, we discuss some of the proposed signatures for the CEP in HICs, concentrating on the possible role played by the finiteness of the system created in these collisions, as proposed in Refs. \cite{Palhares:2009tf,FSS-RHIC}.

\section{The CEP: thermodynamic limit {\it vs} HICs}

In the idealized case of an equilibrium system in the thermodynamic limit, a critical endpoint is always associated with unambiguous, sharp signals due to the second order phase transition that characterizes it. Large fluctuations are expected at all length scales and the correlation length diverges, leading to the conformal invariance present in second order phase transitions (cf. e.g. Ref. \cite{goldenfeld}). Direct consequences and signals are then divergences in susceptibilities and other thermodynamic quantities.

In any real system, however, even if we can assume equilibrium, the volume is finite, so that the partition function (and hence all other thermodynamic quantities) is always well-defined and nonsingular. Similarly, the correlation length does not diverge. In this case, the only way to establish the presence of a second order phase transition is to verify that the correlations in the system satisfy a specific behavior as a function of the system size leading to the conformal invariance that characterizes the second order phase transition in the thermodynamic limit. Such specific trend is the well-known Finite-Size Scaling (FSS) \cite{FSS-books}.

The connection between the CEP phenomena in the thermodynamic limit and the asymptotic particles measured in HIC detectors is even more indirect. The medium created in such experiments is an extremely complicated and fastly evolving system that goes through different stages that cannot in general be disentangled in the final observed spectrum. The QCD phase transition is only one of these stages, being followed by an interacting hadronic medium before the freeze-out that ultimately defines the final particle distributions detected. Moreover, it is probable that the phase conversion process occurs actually far from equilibrium, which could change completely the correlations in the system and even manifest features which are not consistent with the ones expected from the phase structure in the thermodynamic limit.

The problem of defining a clear and unambiguous signature for the CEP in HICs is therefore a nontrivial task that should be pursued with caution, always having the various caveats resulting from the specific experimental setup in mind. In what follows, we will disregard completely several of these caveats (e.g. out-of-equilibrium effects, critical slowing down near the CEP \cite{Berdnikov:1999ph}, hadronic medium modification, etc) and concentrate on the role played by finite-size effects. Besides discussing the caveats brought about by the finiteness of the system, we show how the fact that HICs provide data from media of different sizes may actually allow for a complementary alternative signature of the CEP based on FSS \cite{Palhares:2009tf}.

\section{CEP signatures in HICs}

As discussed above, the CEP in a real system manifests as peaks in the cumulants of the distribution of the order parameter as functions of the distance to the criticality in the external parameter space. In this vein, it is natural to try to construct signatures of the CEP in HICs as a non-monotonic behavior in observables connected to the order parameter of the QCD transitions as one passes through or nearby the critical temperature $T_c$ and baryon chemical potential $\mu_{B c}$. In HICs, however, one does not have direct experimental access to neither of the order parameters for the chiral and deconfinement transitions, namely the chiral condensate $\sigma$ and the expectation value of the Polyakov Loop, respectively.
Moreover, even if we assume that the system is equilibrated, direct measurements of $T$ and $\mu_B$ are not available.

The first concrete proposal of signature that tries to address these issues is the one by Stephanov {\it et al} \cite{Stephanov:1998dy}. The idea is based on the assumption that the critical correlations of the chiral order parameter will be transmitted to particles in the final HIC spectra that can effectively couple to the chiral condensate field $\sigma$, such as pions (with interaction $G\sigma\pi^+\pi^-$) and nucleons ($g_N\sigma\overline{N}N$). Within this effective description, the critical contribution to the correlation of the fluctuations in the particle numbers can be written in terms of the correlation length $\xi$ (or the inverse of the $\sigma$ mass), that should drastically increase (or diverge in the thermodynamic limit) at the CEP. The two-particle correlations give (for more details and notation cf. e.g. Ref. \cite{Athanasiou:2010kw}): $\langle\delta n_p\delta n_k\rangle\sim \xi^2$. Thus, a peak in this observable is expected in the vicinity of the CEP. Its height is related to the maximum correlation length, which in turn is limited by various factors: the size of the nuclei overlapping region (in practice, the effective system size can be considerably small, as for example is the case in core-corona scenarios \cite{Werner:2007bf}),  the finite lifetime of the medium, critical slowing down \cite{Berdnikov:1999ph}, etc.

Besides the critical contribution to the particle number fluctuations, there are also different noncritical contributions (e.g. those coming from thermal decays) that form a background to the CEP signature which could consequently be hidden. With the aim of increasing the sensitivity with respect to the critical behavior near the CEP, different observables that couple to higher cumulants of the order parameter have been proposed. Respectively connected to the three- and four-particle correlators, the critical contributions to skewness $\omega_3\sim \xi^{9/2}$ and kurtosis $\omega_4\sim \xi^{7}$ have stronger dependence on the correlation length, yielding a higher chance that these peaks surmount the noncritical background \cite{Stephanov:2008qz}.

Specific dimensionless ratios of these higher cumulants that are believed to be less model-dependent \cite{Athanasiou:2010kw} have been compared to current available HIC data and, very recently, to the analysis of the first run of the RHIC BES program. Up to now no clear indication of the criticality has been observed in this type of signatures. Instead of the expected increase in the fluctuations at lower energies, one has actually found a decreasing trend in the kurtosis observable. Recent work has been put forward that interprets this apparent drop as the behavior in the direction of negative kurtosis, which is claimed to be a signature of the vicinity of the CEP from the crossover side \cite{Stephanov:2011pb}. Nevertheless, this is not a sufficient condition for the CEP and can be a result of other constraints of the medium created in HIC, such as baryon number conservation \cite{Schuster:2009jv}.

\begin{figure}
\center
\begin{minipage}[t]{60mm}
%\framebox[79mm]{\rule[-26mm]{0mm}{52mm}}
% \center
%{\epsfig{file=pseudocorr.eps,width=5cm}}
\includegraphics[width=5.5cm]{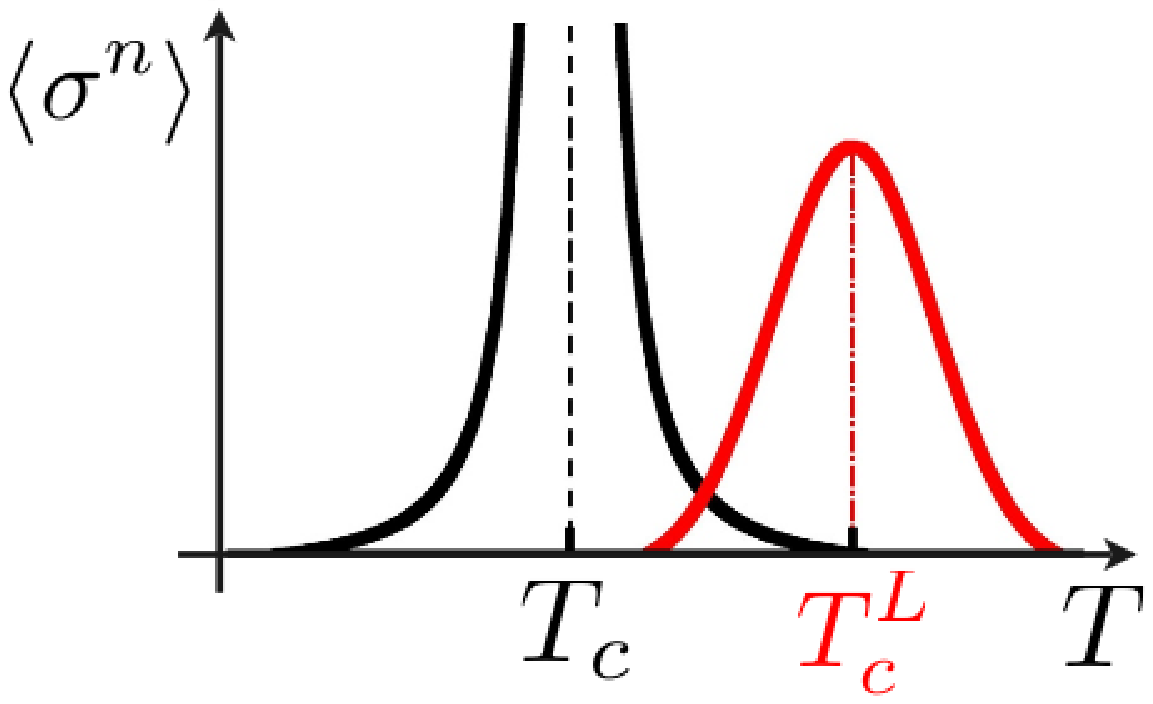}
\noindent\caption{N-point correlation functions of the chiral order parameter $\sigma$ as a function of the external parameter $T$ in the thermodynamic limit (black, left-most line) and for a finite system (red, right-most line).}
\label{pseudo}
\end{minipage}
\hspace{.4cm} 
\begin{minipage}[t]{55mm}
%\framebox[74mm]{\rule[-26mm]{0mm}{52mm}}
% \center
\includegraphics[width=5.3cm]{PhDiagram-PBC.eps}
% \label{Tc-mpi}
\noindent\caption{Pseudo-critical chiral phase diagram for different size scales probed in current HICs as predicted within the Linear Sigma Model coupled to constituent quarks \cite{Palhares:2009tf}.}
\label{PhDiag}
\end{minipage}
\end{figure}

Let us now address the role played by finite-size effects in the case of the signatures mentioned above. As discussed in the previous section, the system created in HICs is finite and this imposes a natural limitation on the growth of the correlation length. Besides this trivial constraint that smoothens out CEP divergences into peaks, the finiteness of the system created in HICs also guarantees that non-monotonic signals will actually probe pseudocritical quantities which might be significantly shifted from the genuine criticality, as illustrated in Fig. \ref{pseudo}. Estimates of these shifts in chiral models show that the displacement of the (pseudo-)CEP and the (pseudo-)first-order lines can be large for the size scales present in current HIC experiments \cite{Palhares:2009tf} (cf. Fig. \ref{PhDiag}). If this is indeed the case, the
non-monotonic behavior signaled by the fluctuation observables will not only be blurred in a region and the effects from criticality severely smoothened, but also this region will be shifted with respect to most of the effective-model and lattice predictions, that are done in general in the thermodynamic limit. Moreover, once finite-size corrections are important, the observables will be also sensitive to boundary effects. Smoothening of the fluctuation peaks will be further increased by the averaging procedure within centrality (i.e. system size) bins, tending to hide them underneath the background. 

\section{Finite-size scaling as a tool for the CEP search}

On the other hand, the conformal invariance expected at the CEP in the thermodynamic limit leads to strong constraints for the behavior of finite systems, so that finite-size effects may, instead of only bringing obstacles for the CEP search, be turned into a tool. The non-monotonic behavior of correlation functions near criticality for systems of different sizes, given by different centralities in HICs, must obey FSS. As described in Ref. \cite{Palhares:2009tf}, it is possible to pragmatically translate the FSS condition to the context of HICs and use the collapsing of curves in an adequate scaling plot as a signal of the CEP.

Ref. \cite{FSS-RHIC} presents a concrete study of the applicability of the predicting power of scaling plots in the search for the CEP of QCD in HICs, using data from RHIC and SPS. The analysis suggests the exclusion of a critical point below chemical potentials $\mu\sim 450 $MeV and, via extrapolations, points to criticality slightly above $\mu\sim 500 $MeV. Assuming FSS, the behavior of new data at lower center-of-mass energies, currently being investigated in the BES, has also been predicted.

\section{Final remarks}

Finite-size effects most probably play an important role in HIC experiments and particularly in the QCD CEP search. As a caveat: restricting the access of non-monotonic signatures to pseudo-critical quantities that are severely smoothened and might be significantly shifted with respect to the genuine CEP (in the thermodynamic limit). But also as a complementary tool: FSS analysis is perfectly feasible and simple to be made in the context of HIC’s, yielding predictions for comparison of data for different center-of-mass energies.

%%%%%%%%%%%%%%%%%%%%%%%%%%%%%%%%%%%%%%%%%%%%%%%
%\section*{Acknowledgments} 
\noindent We are grateful to T. Kodama and P. Sorensen for fruitful collaboration in the topic, to K. Werner for discussions and to the organizers of Excited QCD 2011 for the stimulating meeting. This work was partially supported by CAPES-COFECUB (project 663/10), 
CNPq, FAPERJ and FUJB/UFRJ.

%%%%%%%%%%%%%%%%%%%%%%%%%%%%%%%%%%%%%%%%%%%%%%%

\end{document}